\documentclass[]{aastex63}  

\usepackage{verbatim}
\usepackage{float}
\usepackage{graphicx}
\usepackage{epstopdf}
\usepackage{subfigure}
\usepackage{CJK}

\usepackage{booktabs}
\usepackage{url}
\usepackage{amssymb}
\makeatletter

\newcommand{\Rmnum}[1]{\expandafter\@slowromancap\romannumeral #1@}
\makeatother

\shorttitle{}
\shortauthors{}

\begin{document}
\begin{CJK*}{UTF8}{gbsn}

\title{Gaussian Process Modeling Coronal X-ray Variability of Active Galactic Nuclei}

\author{Haiyun Zhang (张海云)}
\affiliation{Department of Mathematics and Statistics, Yunnan University, Kunming 650091, China; zhanghy@ynu.edu.cn,nstang@ynu.edu.cn}

\author{Dahai Yan (闫大海)}
\affiliation{Department of Astronomy, Key Laboratory of Astroparticle Physics of Yunnan Province, Yunnan University, Kunming 650091, China; yandahai@ynu.edu.cn,lizhang@ynu.edu.cn}

\author{Li Zhang (张力)}
\affiliation{Department of Astronomy, Key Laboratory of Astroparticle Physics of Yunnan Province, Yunnan University, Kunming 650091, China; yandahai@ynu.edu.cn,lizhang@ynu.edu.cn}
\author{Niansheng Tang (唐年胜)}
\affiliation{Department of Mathematics and Statistics, Yunnan University, Kunming 650091, China; zhanghy@ynu.edu.cn,nstang@ynu.edu.cn}

\begin{abstract}
The corona is an integral component of active galactic nuclei (AGNs) which can produce the X-ray emission. 
However, many of its physical properties and the mechanisms powering this emission remain a mystery.
In this work, we study the coronal X-ray variabilities of 13 AGNs by Gaussian Process. 
2-10 keV light curves of 13 AGNs can be successfully described by the damped-random walk (DRW) model. 
The extracted coronal X-ray timescales range from 3 to 50 days.
In the plot of variability timescale versus black hole mass, the coronal X-ray timescales of four sources occupy almost the same region as the optical timescales of the accretion disk, with the latter matching the predicted thermal instability timescale of the disk.
In contrast, the X-ray timescales of the remaining sources exhibit a systematic offset toward lower values.
We propose that the coronal X-ray variability may be driven by internal processes within the corona itself (such as thermal conduction). On the other hand, it may also be triggered by local thermal instabilities occurring in different regions (close to the central black hole) of the accretion disk, which propagate to the corona via disk-corona coupling. 

\end{abstract}   


\section{Introduction} \label{sec:intro}

The X-ray radiation is a common feature of active galactic nuclei (AGNs). 
Different from the X-ray emission of radio loud AGNs, which is mainly contributed by inverse Compton scattering and synchrotron radiation in jets, the continuum intrinsic X-ray emission of radio quiet AGNs is generally believed to be caused by the inverse Compton scattering of thermal optical/ultraviolet photons from the accretion disk by hot electrons in a hot cloud of plasma called the corona \citep[e.g.,][]{2001MNRAS.321..549M}. 
For radio quiet AGNs, there are usually other components or processes, such as reflection component, making X-ray emission complicated \citep[e.g.,][]{2013MNRAS.429.2917F,2021MNRAS.508.1798P}.

Variability occurred in the entire electromagnetic wavelengths is one of the characteristics of AGNs.
The underlying physical process of the variability in the innermost region of AGNs is usually studied by analyzing the X-ray emission from the hot corona \citep[e.g.,][]{2017MNRAS.471.4436W} as well as the UV/optical radiation from the accretion disk \citep[e.g.,][]{2009ApJ...698..895K,2021Sci...373..789B}, which are associated with the emission from the central engine.
Approaches, such as the flux distribution \citep[e.g.,][]{2014MNRAS.440..106B,2017MNRAS.472..174M,2018ApJ...864..164Y},
the power spectral density (PSD) \citep[e.g.,][]{2012AA...544A..80G,2022ApJ...936...36Y}, the spectral variability \citep[e.g.,][] 
{2010LNP...794..203M,2018A&A...619A..93B,2015MNRAS.447.1692Y} and the cross-correlation analysis \citep[e.g.,][]{2017ApJ...840...41E,2019ApJ...870...54M,2020MNRAS.494.4057P}, have been carried out to explore the emission mechanisms and related physical processes.

In addition, a probabilistic statistical analysis method, Gaussian Process (GP), has been applied to analyze the jet and disk variability \citep[e.g.,][]{,2019ApJ...885...12R,2021Sci...373..789B,2021ApJ...919...58Z,2022ApJ...930..157Z,2025MNRAS.537.2380Z,2025MNRAS.tmp..930Z}.
Using this method, the connection between the jet and the accretion disk has been proposed.
To be specific, the long-term variability of jets and optical variability of normal quasars may both be associated with the thermal instability of the accretion disk \citep{2021Sci...373..789B,2019ApJ...885...12R,2022ApJ...930..157Z}.
GP method has also been successfully employed to perform measurement of time lags associated with X-ray reverberation from the accretion disc \citep{2019MNRAS.489.1957W}.

It is generally accepted that the 2-10 keV X-ray emission of quasars is produced in the corona.
In this work, we would like to study the X-ray long-term variability in such a corona structure by GP method, and explore whether there is a correlation among the radiation in corona, jet and accretion disk.  
The published X-ray data from RXTE AGN Timing $\&$ Spectral Database\footnote{\url{https://cass.ucsd.edu/~rxteagn/}} \citep{2013ApJ...772..114R} are used here.
The format of this paper is as follows. 
In Section~\ref{sec:sample}, we describe the samples and GP method. In Section~\ref{sec:results}, we give the modeling results of our samples. The relevant discussion of the results and a summary are given in Section~\ref{sec:discussion} and Section~\ref{sec:summary}, respectively.

\section{Sample and Method} \label{sec:sample}
From the RXTE AGN Timing $\&$ Spectral Database which provided us with 16 yr (1996-2012) of data, we select AGNs that possess abundant data points in their 2-10 keV light curves (LCs) and have black hole masses as low as $10^{6}$-$10^{8} M_{\rm \odot}$.
Blazars are excluded.
Finally, 13 sources are selected, 2 radio galaxies (RDGs) are included. 
Table~\ref{tab:information} gives the general information of these sources.
It is notable that instead of using the whole LCs obtained from the database, we intercept the time series with good sampling rate (without large time intervals as hundreds days), and we list the time range of the analyzed LCs in Table~\ref{tab:fit}.

\begin{deluxetable*}{cccccc}
	\tablecaption{Information of 13 AGNs.\label{tab:information}}
	\tablewidth{0pt}
	\setlength{\tabcolsep}{4mm}{
	\tablehead{
		\colhead{Object} &  \colhead{$z$} & \colhead{Type} & \colhead{$\rm{log}\;M_{\rm BH}/M_{\rm \odot}$} & \colhead{Ref.} & \colhead{Eddington rate}
	}
	\decimalcolnumbers
	\startdata
    3C 120 & 0.033 & RDG & $7.74\pm0.04$ & 1 & 0.44\\
	Mkn 79 & 0.0222 & sey 1.2 & $7.61\pm0.12$ & 1 & 0.07 \\
	NGC 4258 & 0.00149 & sey 2 & 7.6 & 2 & 0.006\\
	3C 111 & 0.049 & RDG & $8.1\pm0.5$ & 1 & $\cdots$\\
	NGC 5548 & 0.0172 & sey 1.5 & $7.91\pm0.18$ & 1 & 0.06\\
	NGC 3227 & 0.0039 & sey 1.5 & $6.77\pm0.1$ & 1 & 0.10\\
	NGC 4051 & 0.0023 & NLS1 & $6.13\pm0.14$ & 1 & 0.23\\
	Mkn 766 & 0.0129 & NLS1 & $6.6$ & 2 & 0.15\\
	NGC 4593 & 0.009 & sey 1 & $6.88\pm0.09$ & 1 & 0.20\\
	NGC 7213 & 0.00584 & sey 1.5 & 7.7 & 2 & 0.03\\
	NGC 7469 & 0.01627 & sey 1.5 & $6.98\pm0.05$ & 1 &0.32\\
    NGC 3783 & 0.00973 & sey 1.5 & $7.47\pm0.07$ & 1 & 0.07\\
    NGC 5506 & 0.00608 & NLS1 & 7.4 & 2 &0.03\\ 
	\enddata 
	\tablecomments{(1) source name, (2) redshift, (3) source type, (4) black hole mass (in solar mass) collected from the references in the (5) column, (6) Eddington rate, the ratio between the total luminosity $L_{\rm bol}$ and a mass-dependent characteristic luminosity $ L_{\rm Edd}=1.3\times 10^{38}(M_{\rm BH}/M_{\rm \odot})\;erg\;s^{-1}$, $L_{\rm bol}$ is from\cite{2012AA...544A..80G}}. The $L_{\rm bol}$ of Mrk 79 is calculated through $L_{\rm bol}/L_{\rm 5100}=9.26$, $L_{\rm 5100}$ is from the AGN Black Hole Mass Database Web\footnote{\url{http://astro.gsu.edu/AGNmass/details.php?varname=7}}.
    References: 1 \cite{2021Sci...373..789B}, 
 2 \cite{2012AA...544A..80G}.}
\end{deluxetable*}

We characterize the X-ray LCs by GP method.
GP is a class of statistical models built on stochastic processes and provides a powerful and flexible tool for processing astronomical data.
A mean function and a covariance function (kernel) can totally determine a GP.
The mean function is set to be the mean flux of the LC.
The kernel we use here is the damped-random walk (DRW) model implemented in {\it celerite} package \citep{2017AJ....154..220F}.
A detailed description of GP method, DRW model, and data processing processes can be found in our previous work \citep{2022ApJ...930..157Z,2023ApJ...944..103Z}, and we will give a concise introduction here.
The DRW is defined by an exponential covariance function:
\begin{equation}\label{eq1}
k(t_{nm})=2\sigma_{\rm DRW}^{2}\cdot \exp(-t_{nm}/\tau_{\rm DRW})\;\ ,
\end{equation}
where $t_{\rm nm} = |t_{\rm n}-t_{\rm m}|$ is the time lag between measurements $m$ and $n$.
$\sigma_{\rm DRW}$ represents the amplitude term, and $\tau_{\rm DRW}$ represents the damping timescale, describing the rate at which the correlation of the data decays over time.
DRW has PSDs in the form of a broken power-law with the spectrum index being 0 at lower frequencies and -2 at higher frequencies.
When the DRW model fails to adequately describe the observed LCs, we add an excess white noise term ($\sigma_{n}\delta_{nm}$) to the model, which is referred to as the DRW(3) model. 
The original DRW model is termed the DRW(2) model. 

The total fitting process using {\it celerite} package is as follows. We first assume log-uniform priors on each of the DRW parameters and determine the initial values for the Markov Chain Monte Carlo (MCMC) sampling using a maximum-likelihood estimate. Then we run 50,000 MCMC iterations, discarding the first 20,000 as burn-in. 
The remaining 30,000 fits provide the corresponding parameter sets and the PSD values at every frequency. They are respectively used to plot the posterior distributions of the parameters and to construct the final PSD. 

In this study, we assess the goodness of fit by comparing the model predictions to the observational data, evaluating the distribution of standardized residuals to test for normality, and analyzing their autocorrelation function (ACF) to check for consistency with white noise within the 95\% confidence interval  \citep{2021ApJ...907..105Y}.
The derived damping timescale is reliable if the following constraints are satisfied  \citep{2021ApJ...907...96S,2021Sci...373..789B}:\\
\indent(1) $\tau_{\rm DRW}$-Length: $\tau_{\rm DRW}\;\textless\;0.1\times$ the length of LC;\\
\indent(2) Sampling: $\tau_{\rm DRW}\;\textgreater\;$ cadence, where cadence is the mean cadence of the LC; \\
\indent(3) AGN-like variability: P$\textless$0.05, where P is the $p$-value of Ljung-Box test\citep{10.1093/biomet/65.2.297}. This constraint rejects the hypothesis that the LC is white noise.\\

\begin{deluxetable*}{ccccccc}
	\tablecaption{Modeling information for 13 AGNs.\label{tab:fit}}
	\tablewidth{50pt}
	\setlength{\tabcolsep}{2mm}{
	\tablehead{
		\colhead{Name} &\colhead{Model} &\colhead{Time} &\colhead{Cadence} & \colhead{$p$ value}  \\
		\colhead{} & \colhead{} & \colhead{((MJD))} & \colhead{(days)} & \colhead{} \\
		\colhead{(1)} & \colhead{(2)} & \colhead{(3)} & \colhead{(4)} & \colhead{(5)}
	}
	\startdata
	3C 120 & DRW(2) &52334.9-54221.9 & 1.65 & 0.51 \\
	Mkn 79 & DRW(3) & 52720.4-55925.6 & 1.99 & 0.003 \\
	NGC 4258 & DRW(2) & 53436.6-55924.1 & 3.13 & 0.23\\
	3C 111 & DRW(2) & 53065.4-55924.4 & 2.94 & 0.40\\
	NGC 5548 & DRW(2) & 50208.0-52749.7 & 3.41 & 0.14\\
	NGC 3227 & DRW(3) & 51180.5-53709.0 & 2.83 & 0.11\\
	NGC 4051 & DRW(3) & 50196.5-55925.5 & 2.70 & $3.6\times10^{-5}$\\
	Mkn 766 & DRW(3) & 53065.4-55924.1 & 3.80 & 0.03 \\
	NGC 4593 & DRW(3) & 53063.4-55926.6 & 2.24 & 0.56\\
	NGC 7213 & DRW(2) & 53797.1-55195.0 & 1.71 & 0.44 \\
	NGC 7469 & DRW(2) & 53470.8-53777.9 & 4.39 & 0.94 \\
	NGC 3783 & DRW(3) & 53063.4-55928.4 & 2.21 & 0.13 \\
	NGC 5506 & DRW(3) & 50159.8-52414.5 & 3.58 & 0.05
	\enddata
	\tablecomments{ 
	(1) source name, (2) model, (3) time period of the analyzed LCs, (4) the mean cadence of the LCs, (5) $p$-value of KS test on the normality of the standardized residuals for the fit.  
		}}
\end{deluxetable*}

\section{results}\label{sec:results}
\subsection{DRW fitting results}
The P values of Ljung-Box test for all LCs are less than 0.05. 
We fit  LCs of 13 AGNs with DRW model, and list their fitting information and parameters in Table~\ref{tab:fit} and Table~\ref{tab:Posterior Parameters}, respectively. 
Whether we use the DRW(2) or DRW(3) model is completely determined by the characteristics of the data themselves rather than subjective preference.
For NGC 4051, Mkn 766 and Mkn 79, although the distributions of standardized residuals deviate from the standard normal distribution (with $p\textless 0.05$ in KS test), they fit normal distributions
with mean values of $-0.06\pm0.03$,$-0.08\pm0.06$, and $-0.04\pm0.03$ respectively, and the same standard deviation of $0.9\pm0.03$. 
We conclude that their LC structures are still adequately captured by the DRW(3) model.
The X-ray LCs of all selected sources are well described by the DRW model, yielding damping timescales of roughly 3-50 days. (listed in Table~\ref{tab:Posterior Parameters}).
 
\begin{figure}
    \centering
    {\includegraphics[width=8cm]{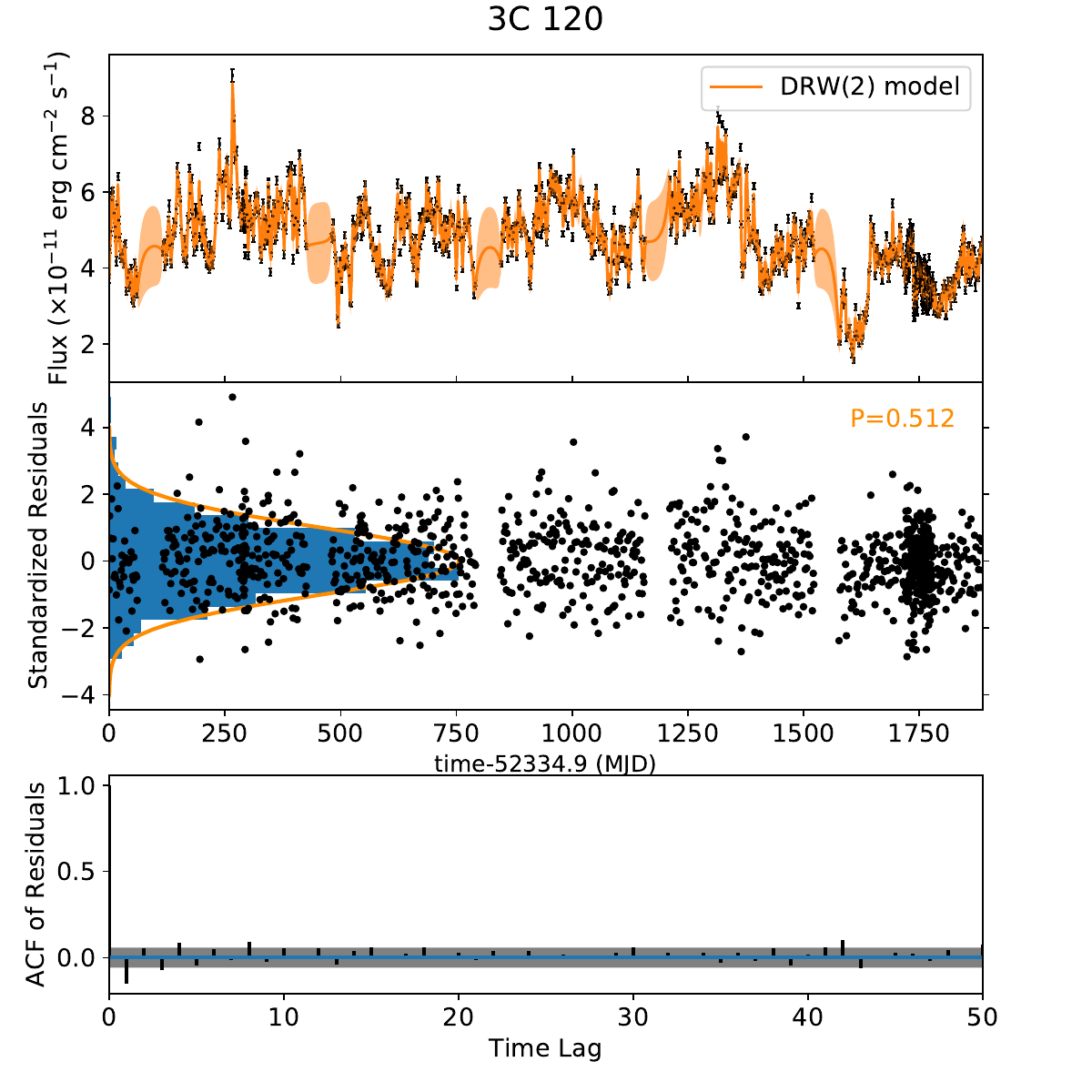}}
    {\includegraphics[width=8cm]{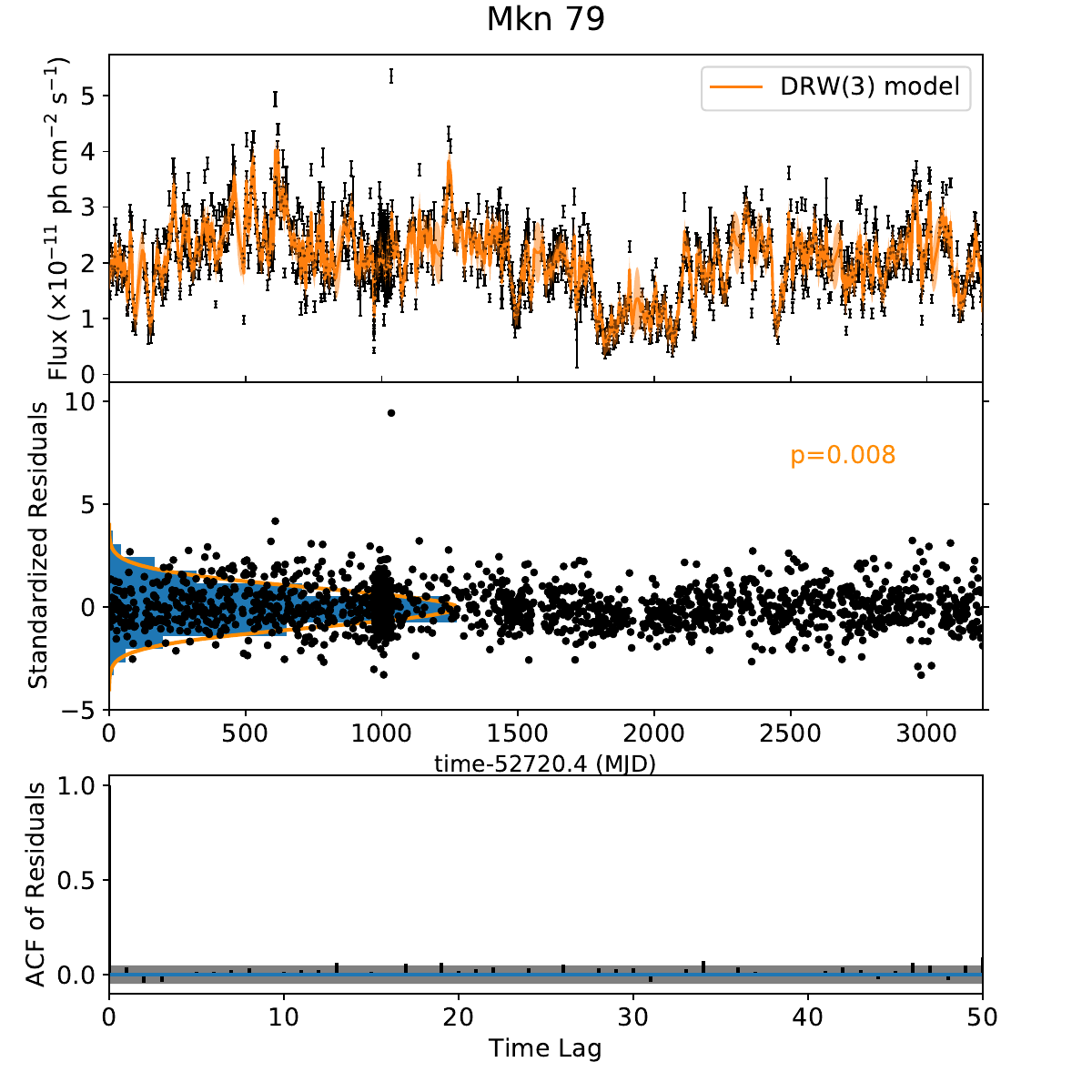}}
    \caption{Fitting results of 3C 120 and Mrk 79 for example. 
    3C 120 is modeled by DRW(2), and Mrk 79 is modeled by DRW(3). 
    For each source, the top panel presents the X-ray LC (black points) and the modeled LC (orange line). 
     We show the standardized residuals (black points), the probability density of standardized residuals (blue histogram) as well as the best-fit normal distribution (orange solid line) in the middle panel. The p value of KS test is labeled in the figure. 
    The ACF of residuals with the 95$\%$ confidence limits of the white noise (the gray region) are shown in the bottom panel.
\label{fig:celerite fit}}
\end{figure}

\begin{figure}
    \centering
    {\includegraphics[width=8cm]{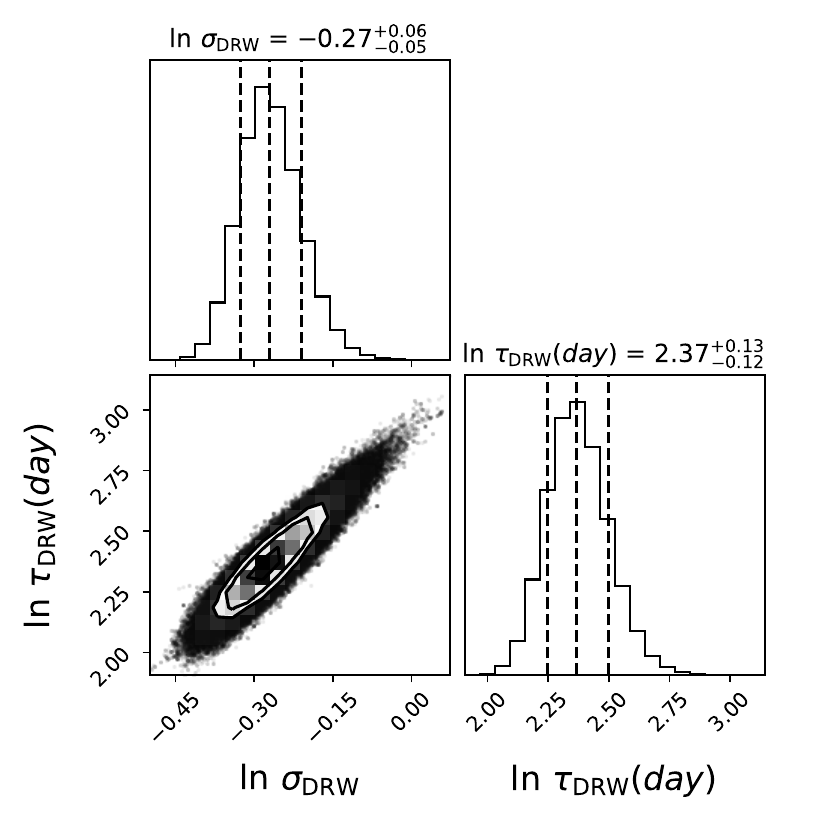}} 
    {\includegraphics[width=8cm]{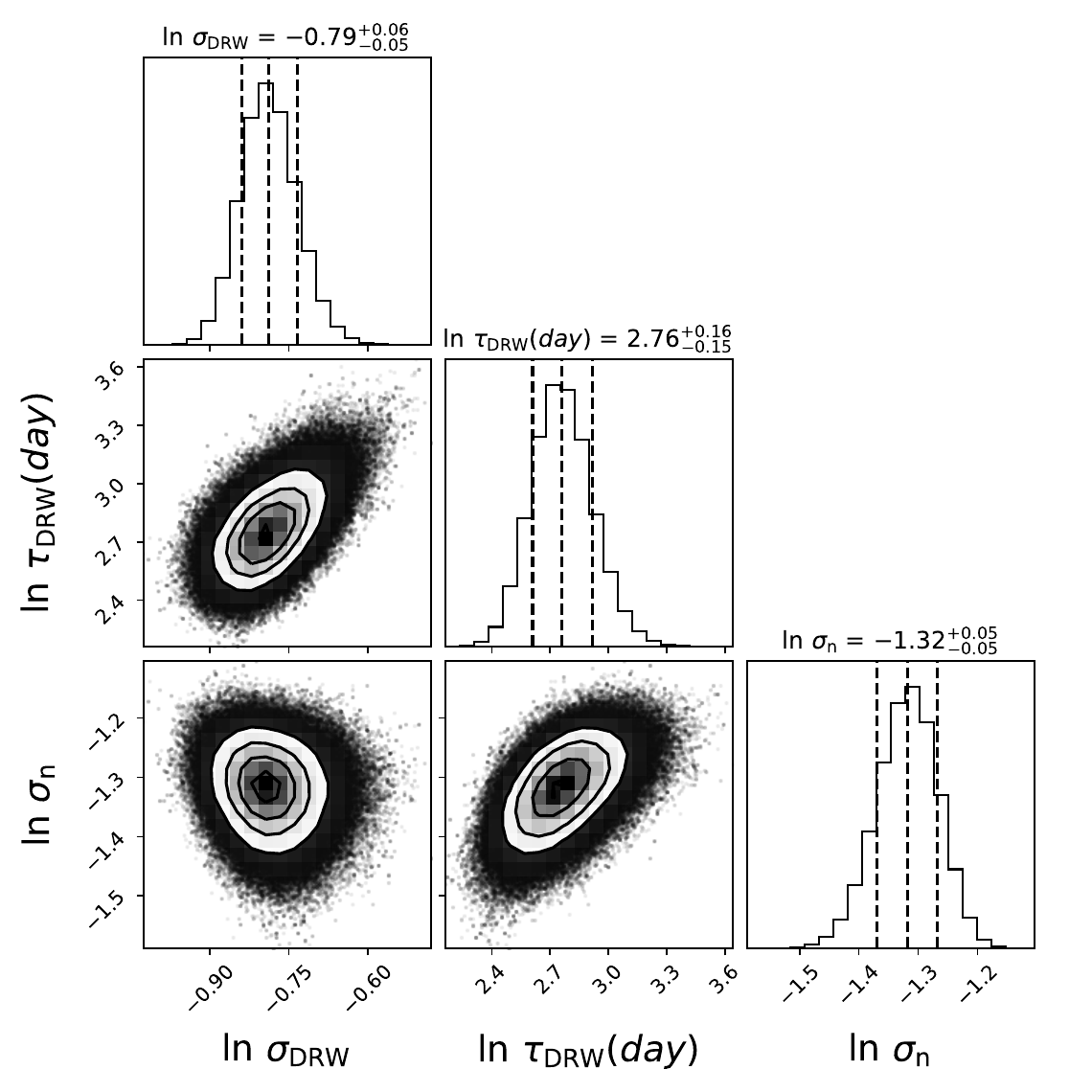}}
\caption{The posterior probability densities of model parameters for 3C 120 (DRW(2) on the left panel) and Mkn 79 (DRW(3) on the right panel). The vertical dotted lines represent the median value and
68$\%$ confidence intervals of the distribution of the parameter.
\label{fig:param distribution} }
\end{figure}

The fitting graphs (Figure~\ref{fig:celerite fit}), posterior distribution of parameters (Figure~\ref{fig:param distribution}) 
of 3C 120 and Mkn 79 are depicted as examples to show the fitting results of DRW(2) and DRW(3) models.
They show strong goodness-of-fit for modeling the LCs and robust constraints for parameters. The PSDs for all sources are in broken power-law form with indexes transiting from -2 at higher frequencies to 0 at lower frequencies.

\begin{deluxetable*}{cccccccc}
	\tablecaption{Posterior Parameters for 13 AGNs.\label{tab:Posterior Parameters}}
	\tablewidth{50pt}
	\setlength{\tabcolsep}{2mm}{
	\tablehead{
    		\colhead{Name} &\colhead{Model}  &  \multicolumn{2}{c}{Parameter of DRW(2)} &  \multicolumn{3}{c}{Parameter of DRW(3)} &\colhead{Damping timescale$^{*}$}  \\
		\cmidrule(r){3-4}\cmidrule(r){5-7}
		\colhead{} & \colhead{} & \colhead{ln $\sigma_{\rm DRW}$} & \colhead{ln $\tau_{\rm DRW}^{*}$}  & \colhead{ln $\sigma_{\rm DRW}$} & \colhead{ln $\tau_{\rm DRW}^{*}$} & \colhead{ln $\sigma_{\rm n}$} & \colhead{(uncorrected)} \\
		\colhead{(1)} & \colhead{(2)} & \colhead{(3)} & \colhead{(4)} & \colhead{(5)} & \colhead{(6)} & \colhead{(7)} & \colhead{(8)} 
	}
	\startdata
	3C 120 & DRW(2) & $-0.27^{+0.06}_{-0.05}$ & $2.37^{+0.13}_{-0.12}$ & $\cdots$ & $\cdots$ & $\cdots$ & $10.7^{+1.4}_{-1.3}$ \\
	Mkn 79 & DRW(3) & $\cdots$ & $\cdots$ & $-0.79^{+0.06}_{-0.05}$ & $2.76^{+0.16}_{-0.15}$ & $-1.32^{+0.05}_{-0.05}$ & $15.8^{+2.5}_{-2.4}$ \\
	NGC 4258 & DRW(2) & $-2.05^{+0.05}_{-0.05}$ & $2.46^{+0.15}_{-0.14}$ & $\cdots$ & $\cdots$ & $\cdots$ & $11.7^{+1.8}_{-1.6}$ \\
	3C 111 & DRW(2) & $0.03^{+0.08}_{-0.07}$ & $3.42^{+0.17}_{-0.15}$ & $\cdots$ & $\cdots$ & $\cdots$ & $30.6^{+5.2}_{-4.6}$ \\
	NGC 5548 & DRW(2) & $0.32^{+0.07}_{-0.06}$ & $3.11^{+0.16}_{-0.14}$ & $\cdots$ & $\cdots$ & $\cdots$ & $22.4^{+3.6}_{-3.1}$ \\
	NGC 3227 & DRW(3) & $\cdots$ & $\cdots$ & $0.2^{+0.09}_{-0.08}$ & $3.36^{+0.28}_{-0.26}$ & $-0.21^{+0.05}_{-0.06}$ & $28.8^{+8.1}_{-7.5}$ \\
	NGC 4051 & DRW(3) & $\cdots$ & $\cdots$ & $-0.36^{+0.07}_{-0.07}$ & $3.61^{+0.23}_{-0.21}$ & $0.01^{+0.02}_{-0.02}$ & $37.0^{+8.5}_{-7.8}$ \\
	Mkn 766 & DRW(3) & $\cdots$ & $\cdots$ & $-0.78^{+0.11}_{-0.10}$ & $3.83^{+0.34}_{-0.30}$ & $-0.53^{+0.04}_{-0.04}$ & $46.1^{+15.7}_{-13.8}$ \\
	NGC 4593 & DRW(3) & $\cdots$ & $\cdots$ & $-0.06^{+0.04}_{-0.04}$ & $1.79^{+0.15}_{-0.14}$ & $-0.73^{+0.10}_{-0.11}$ & $6.0^{+0.9}_{-0.8}$ \\
	NGC 7213 & DRW(2) & $-0.94^{+0.17}_{-0.13}$ & $3.96^{+0.36}_{-0.27}$ & $\cdots$ & $\cdots$ & $\cdots$ & $52.46^{+18.89}_{-14.16}$ \\
	NGC 7469 & DRW(2)  & $-0.57^{+0.11}_{-0.10}$ & $1.71^{+0.31}_{-0.23}$ & $\cdots$ & $\cdots$ & $\cdots$ & $5.53^{+1.71}_{-1.27}$ \\
	NGC 3783 & DRW(3) & $\cdots$ & $\cdots$ & $0.14^{+0.05}_{-0.05}$ & $2.19^{+0.17}_{-0.15}$ & $-0.18^{+0.06}_{-0.06}$ & $8.94^{+1.52}_{-1.34}$ \\
	NGC 5506 & DRW(3) & $\cdots$ & $\cdots$ & $0.33^{+0.07}_{-0.07}$ & $1.56^{+0.31}_{-0.28}$ & $0.23^{+0.07}_{-0.08}$ & $4.76^{+1.48}_{-1.33}$ \\
	\enddata
	\tablecomments{ 
	(1) source name, (2) model, (3)-(4),(5)-(7) posterior parameters of modeling  LCs with DRW(2) model and DRW(3) model respectively, (8) damping timescale. The uncertainties of model parameters and damping timescales represent $1\sigma$ confidence intervals. 
	$^{*}$ indicates that the unit is days.
		}}
\end{deluxetable*}

\subsection{Origin of the 2-10 keV X-ray Emission from 3C 120 and 3C 111}
3C 120 and 3C 111 are radio loud AGNs which are different from other sources.
They both have single side jet.
The presence of jet makes it indeterminate for the location of the X-ray radiation.
The X-ray spectrum of 3C 120 revealed the presence of a Compton reflection component and a moderately broad Fe K$\alpha$ line, indicating that there is X-ray emission from the corona, some of which are scattered and interact with the accretion disk in close proximity to the central black hole \citep{2001ApJ...551..186Z,2012ApJ...752L..21C}.
\cite{2009ApJ...696..601M} also suggested that the majority of the X-ray emission in 3C 120 arises in the disk or corona, but not in the jet.

3C 111 has similar X-ray emission. 
Weak Fe K$\alpha$ line near 6.4 keV was also detected. 
The analysis of the composite spectrum of 3C 111 suggested that although there is contribution of the jet, the thermal part is dominating in the X-ray spectrum during observational period, i.e., the X-ray is mostly of thermal inverse Compton origin \citep{2011ApJ...734...43C,2012A&A...545A..90D}.
We therefore believe that the X-ray emission we analyzed for all sources is predominantly from the coronal region.

\subsection{Damping timescale-Black hole mass Relation} \label{subsec:tau-mass}
In our previous work, we established a possible connection between the jet and the accretion disk radiation based on the plot of variability timescale versus black hole mass \citep{2022ApJ...930..157Z,2023ApJ...944..103Z}.
Here, we include the X-ray characteristic timescales derived from coronal variability in the plot to investigate the origin of variability across different AGN structures.

\begin{figure}
    \centering
       {\includegraphics[width=15cm]{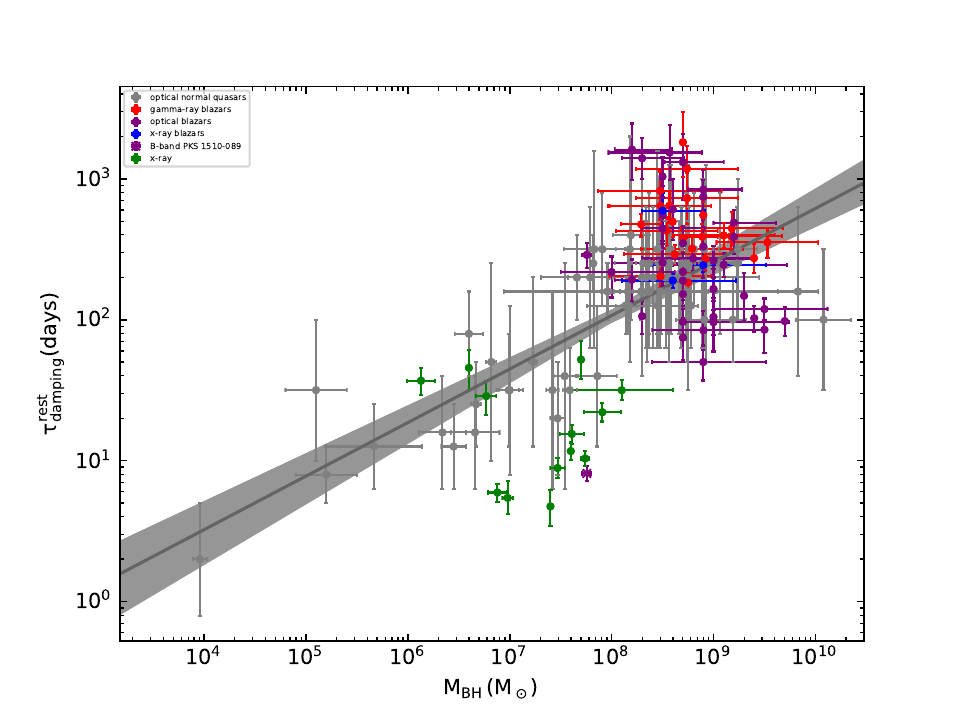}}
  \caption{The plot of rest-frame damping timescale versus black hole mass for AGNs. The green data points represent results of our sample. Red, purple and blue data points represent $\gamma$-ray, optical and X-ray results obtained from \cite{2022ApJ...930..157Z} and \cite{2023ApJ...944..103Z}, and the gray data points and line which represent optical results of accretion disk are from \cite{2021Sci...373..789B}.
    \label{fig:tau-fit}}
\end{figure}

To obtain the rest-frame timescale, it is necessary to adjust the observed timescale for redshift:
\begin{equation}\label{eq4}
\tau_{\rm dampling}^{\rm rest}=\tau_{\rm DRW}/(1+z)\;.
\end{equation} 
In Figure~\ref{fig:tau-fit}, we present the correlation between the intrinsic X-ray timescales of AGNs obtained in this study and black hole masses, along with the multi-waveband jet results reported by \cite{2023ApJ...944..103Z} and optical accretion disk findings from \cite{2021Sci...373..789B}.
We expand the analysis to a lower mass range ($10^{6}$-$10^{8} M_{\rm \odot}$), which is two orders of magnitude smaller than those samples we applied when analyzing jet variability.

Statistically, in such a black hole mass range, our X-ray variability timescales $\tau_{\rm dampling}^{\rm rest}$ range from several days to dozens of days with smaller errors. 
In Figure~\ref{fig:tau-fit}, four sources (NGC 7213, NGC 4051, Mkn 766 and NGC 3227) exhibit coronal X-ray timescales comparable to the optical timescales of the accretion disk, while the remaining nine sources show timescales that systematically fall below the optical regime.

\subsection{Impact of rapid variability} \label{subsec:simulate}
It is notble that RXTE data with cadence of 1-4 days may miss some rapid variability. 
To assess the influence of rapid variability on our long-term variability measurements, we performed simulations and model fitting of the data.
We generate synthetic LCs whose true variability are two superimposed DRW processes representing distinct variability timescales:
(1) a long-term variability component with $\tau_{1}=10$ days and $\sigma_{1}=0.8$; (2) a short-term variability component with $\tau_{2}=0.3$ days and $\sigma_{2}=n\times0.8$ ($n=\sigma_{2}/\sigma_{1}$).
This demonstrates that the original datasets incorporate both long-term variability and rapid fluctuation components. 

\begin{figure}
    \centering
       {\includegraphics[width=0.45\textwidth]{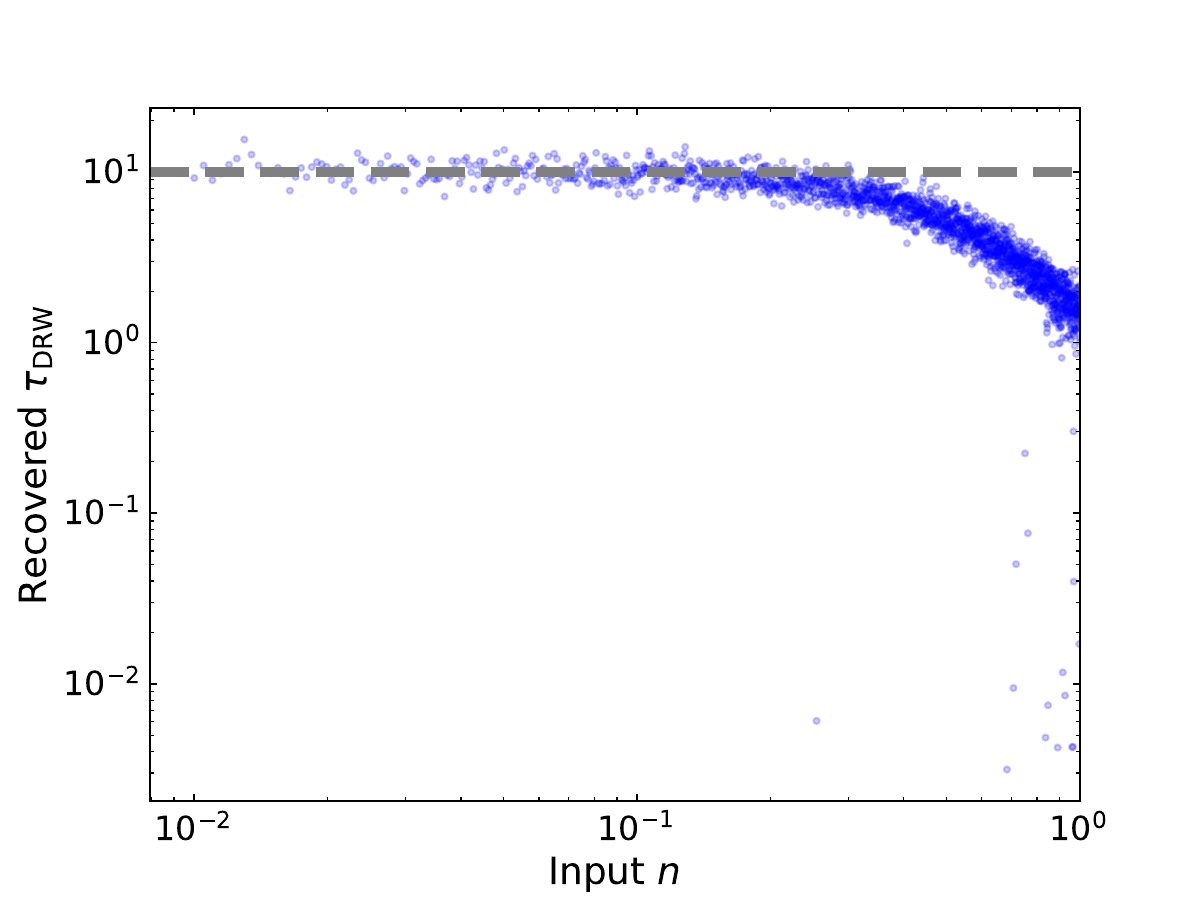}}
       {\includegraphics[width=0.45\textwidth]{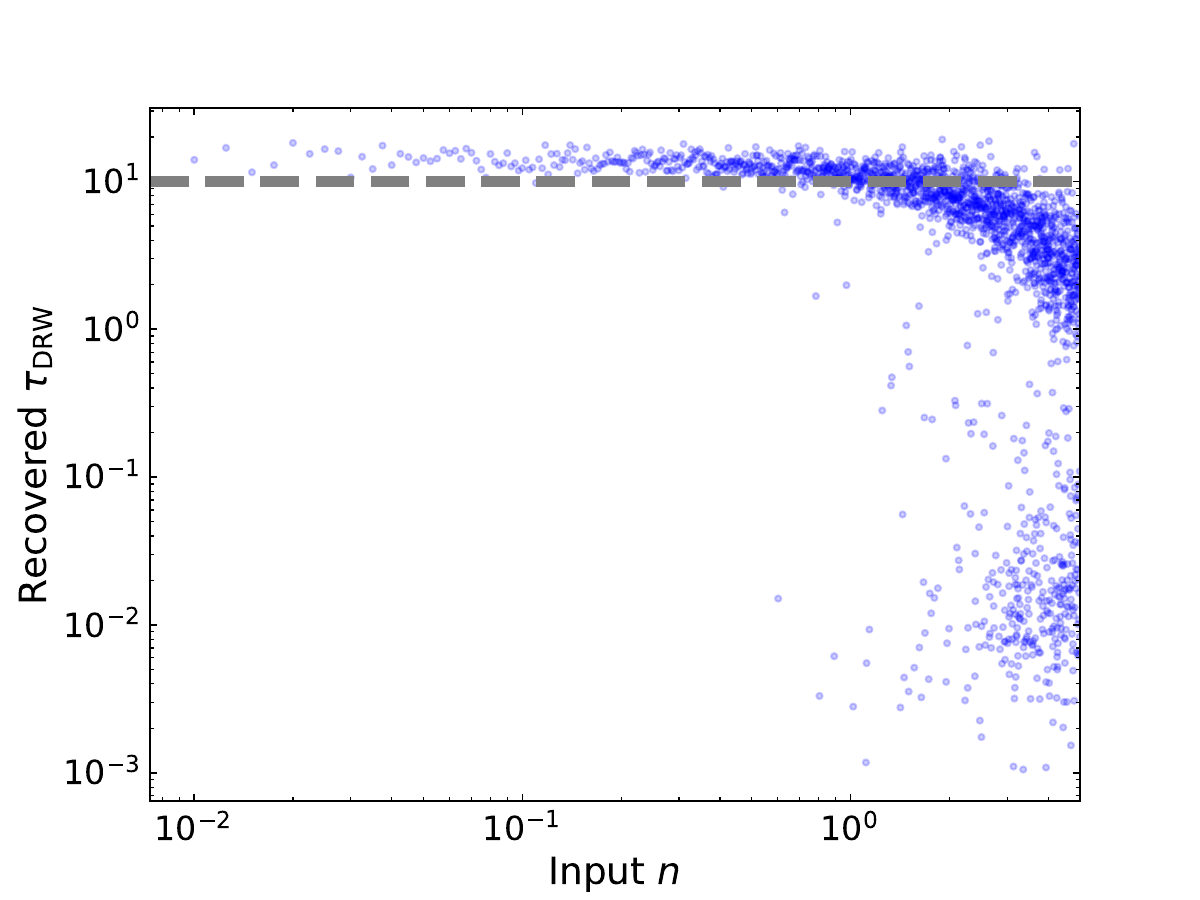}}
  \caption{The variation of the recovered $\tau_{\rm DRW}$ values obtained from fitting simulated LCs with DRW(2) (the left panel) and DRW(3) (the right panel) models as $\sigma_{2}$/$\sigma_{1}$ varies. 
    \label{fig:tau-simulation}}
    \end{figure}

We first generate LCs of total length of 2000 days with non-uniform sampling and an average cadence of 0.05 days, and then rebin these to non-uniform LCs with an average cadence of 2 days, mimicking telescope time resolution and observing patterns.
During the rebinning process, we employ inverse-variance weighted averaging to combine multiple observations.
Each rebinned LC is then fit with a single DRW(2) or DRW(3) model to recover its characteristic timescale. 
The parameter $n$ is evenly spaced and uniformly sampled between 0.01 and 1(5), with a total of 2000 points.
The simulation results are shown in Figure~\ref{fig:tau-simulation}.
When $n<0.2$, DRW(2) model can accurately recover the input $\tau_{1}$ value. When $n$ falls between 0.2-0.3, DRW(2) model recovers $\tau_{1}$ with relatively small deviation (slight underestimation).
While for DRW(3) model, when $n<1$, DRW(3) model exhibits a systematic overestimation of $\tau_{1}$ to some extent, otherwise it underestimates $\tau_{1}$.  
The range of $1<n<2$ represents a transitional regime. 

\begin{figure}
    \centering
       {\includegraphics[width=0.45\textwidth]{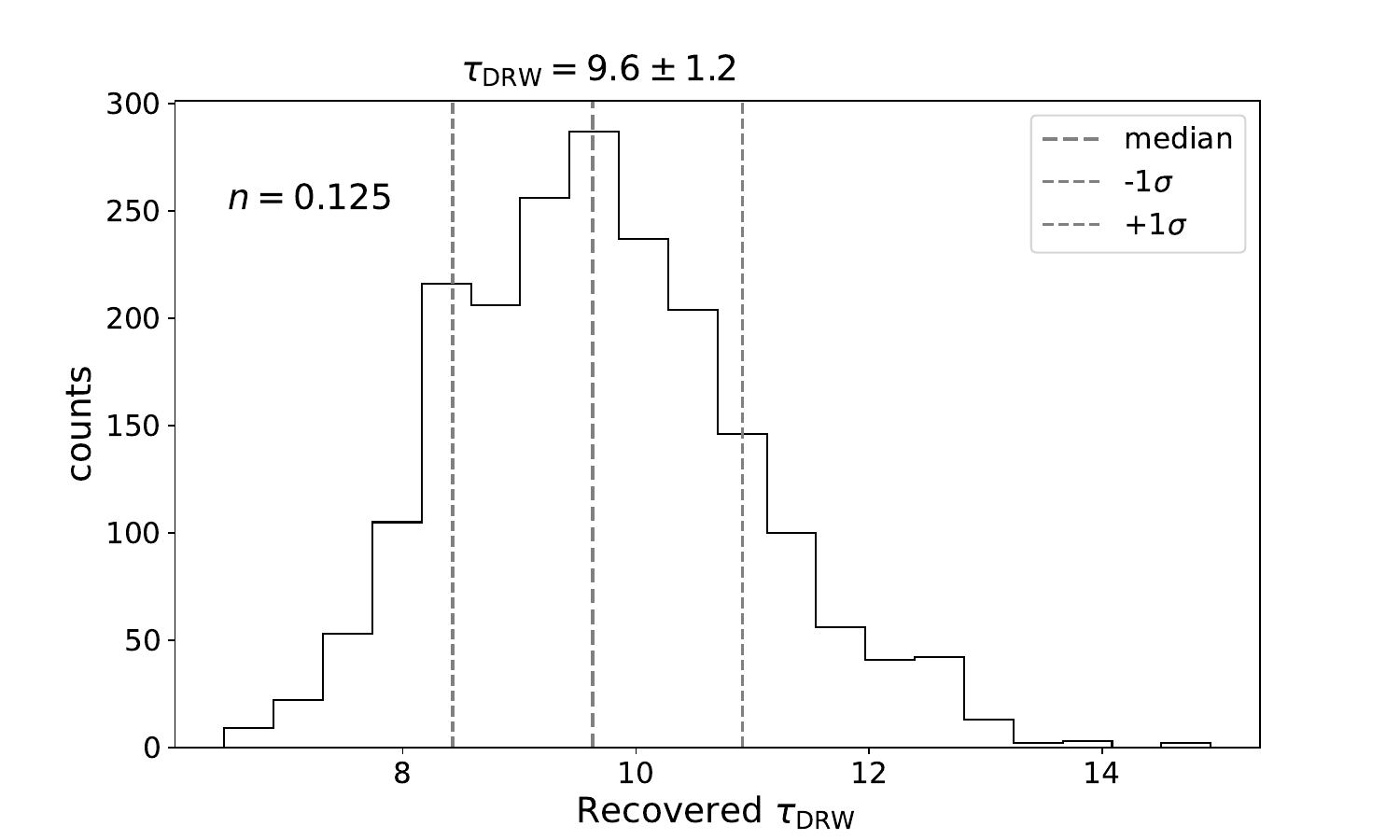}}
       {\includegraphics[width=0.45\textwidth]{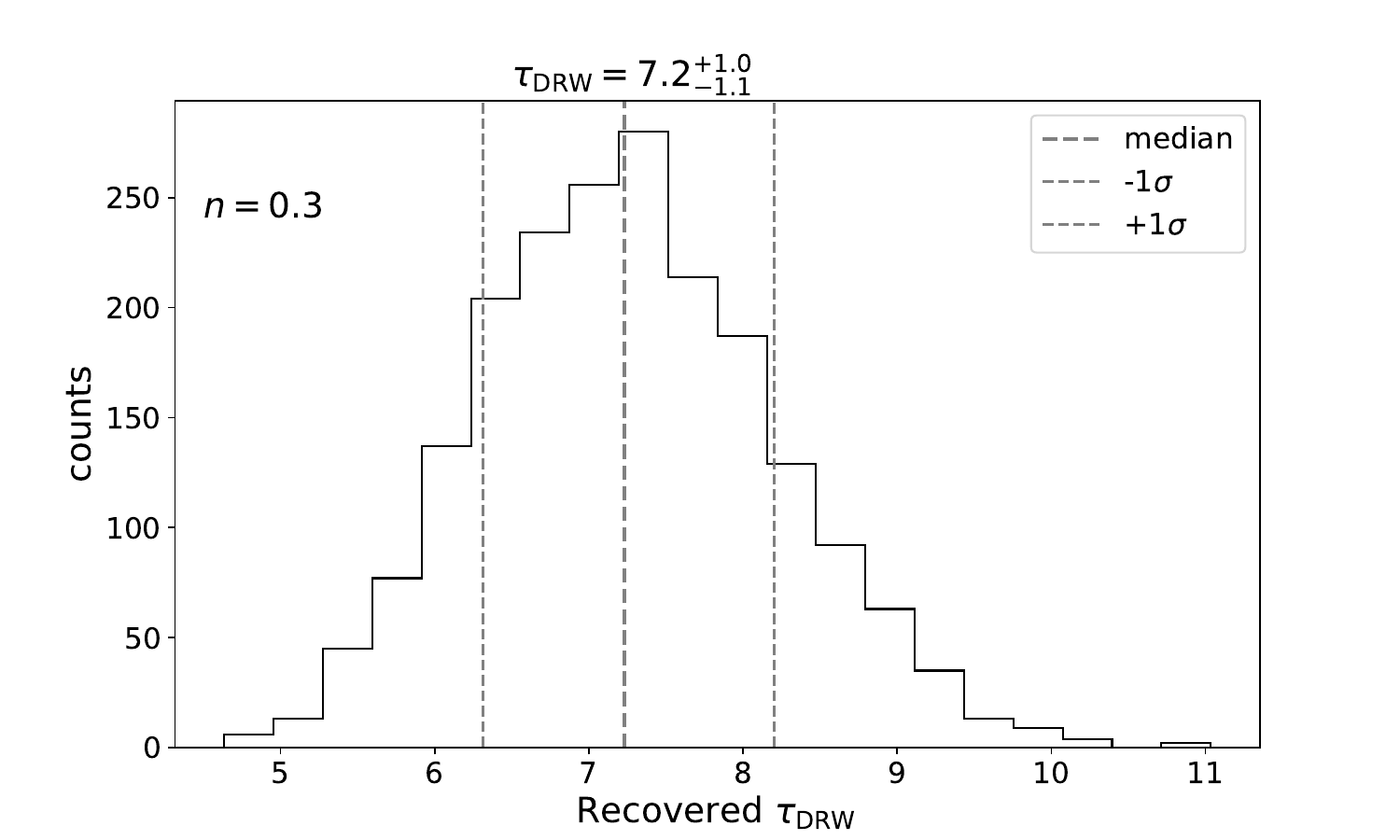}}
       {\includegraphics[width=0.45\textwidth]{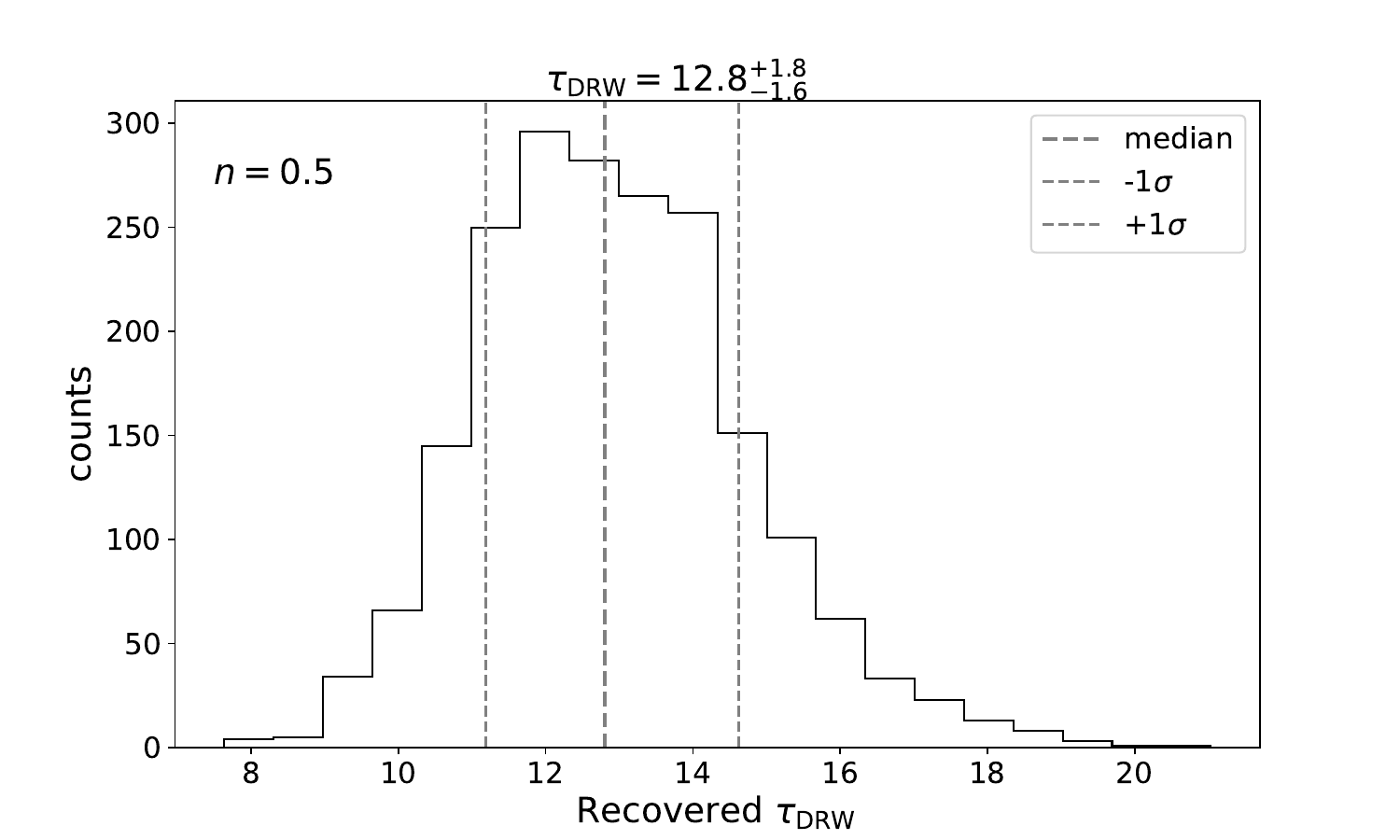}}
       {\includegraphics[width=0.45\textwidth]{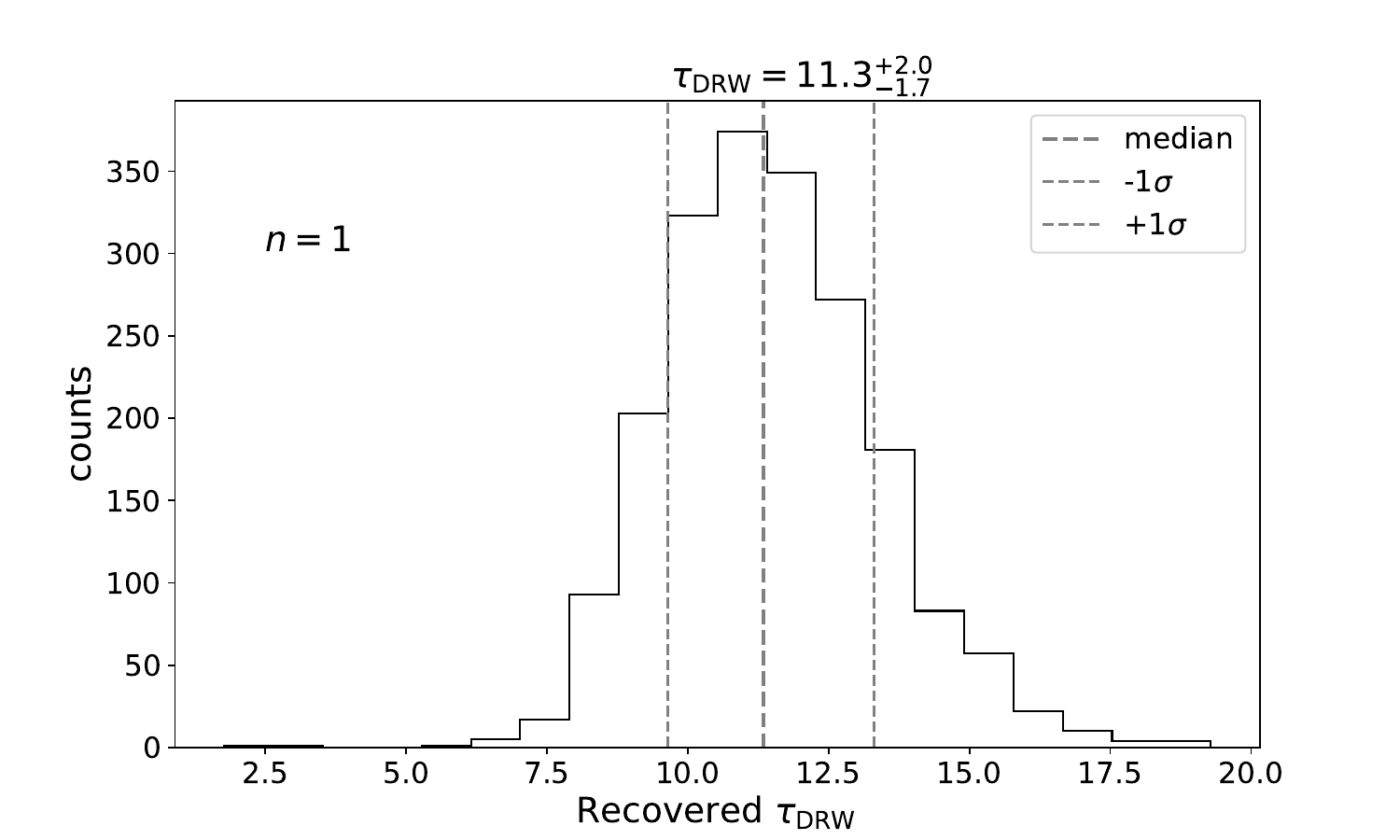}}
  \caption{Distributions of the recovered damping timescales obtained by fitting the simulated data with DRW(2) (upper panels) and DRW(3) (lower panels) with different value of $n$.
    \label{fig:tau-recover}}
    \end{figure}

To quantitatively evaluate the deviation, we assign distinct values to $n$ and perform fitting procedures using both the DRW(2) and DRW(3) models, repeating this fitting process 2000 times for each configuration.
We start from $n=0.125$, the resulting distribution of DRW(2) recovered $\tau_{\rm DRW}$ values (shown in the upper left panel of Figure~\ref{fig:tau-recover}) has a median of 9.6 days with a 1$\sigma$ uncertainty of $\pm$ 1.2 days, which is consistent with the input value.
As the value of $n$ increases (particularly when $n > 0.2$), the deviation between the DRW(2) recovered $\tau_{\rm DRW}$ and input $\tau_{1}$ gradually increases. 
When $n=0.3$, DRW(2) model underestimates $\tau_{1}$ with $\sim 30\%$ (the upper right panel of Figure~\ref{fig:tau-recover}).
When $n=0.5$, DRW(3) model overestimates $\tau_{1}$ with $\sim 30\%$ (the lower left panel of Figure~\ref{fig:tau-recover}), and it overestimates $\tau_{1}$ with $\sim 11\%$ (the lower right panel of Figure~\ref{fig:tau-recover}) when $n$ approaches 1.
In practice, we typically compare the fitting results of distinct models to select the better-performing one. Consequently, the DRW model may introduce biases approximately $0-30\%$ due to the influence of rapid variability components.
Such deviations may arise because, once the high-frequency component becomes comparable to the low-frequency component, a single DRW model will trend to accommodate the combined autocorrelation effects of two distinct timescales to minimize the overall residuals during the fitting, resulting in a shift of the inferred damping timescale.

The 1-4 day observational cadence of the RXTE data we used cannot resolve variability timescales shorter than the sampling interval. Ultimately, the rapid variability component will manifest as a jitter term (white noise). That is, the additional white noise added in the DRW(3) model can be interpreted as originating from the potential rapid variability components of the system itself.

Even after accounting for the impact of potential rapid variability on our results, the overall timescales still roughly fall within the range of 3-50 days, and there is no large influence for the regions of coronal X-ray timescales occupied in Figure~\ref{fig:tau-fit}.
Note that the results for NGC 4051 suggest that shorter-timescale variability components may dominate with $\sigma_{\rm n} > \sigma_{\rm DRW}$, but its recovered $\tau_{\rm DRW}$ has small bias, as $\sigma_{\rm n} / \sigma_{\rm DRW}=1.4$ being in the transitional regime of DRW(3) model.

\section{Discussion} \label{sec:discussion}

The variability in accretion disk of AGNs has been extensively analyzed by stochastic process method including GP, and we have also conducted studies on the long-term variability in jet of AGNs within the black hole mass range of $10^{8}$-$10^{10} M_{\rm \odot}$ by GP method \citep{2022ApJ...930..157Z,2023ApJ...944..103Z}.
In this work, we consider the variability in another basic structure of AGN, namely the corona, and expand the black hole mass of the sample to a lower range. 

The 2-10 keV RXTE LCs of 13 sources in our sample have been characterized by GP method in this study. They are well described by the DRW model, a variability mode that also characterizes the long-term variability observed in both the accretion disk and the jet \citep{2021Sci...373..789B,2023ApJ...944..103Z}.
In our sample, more than half of the sources require an additional jitter term (white noise) in the DRW model to properly describe their X-ray variability. For this jitter term, we attribute it not to systematic errors but rather to rapid DRW variability with short damping timescales.

The coronal X-ray timescales corresponding to the PSD break fall within the range of 3-50 days.
Adding these timescales to the $\rm \tau_{damping}^{rest}-M_{BH}$ plot, we found that the derived corona X-ray timescales have smaller errors and occupy a systemically lower region compared to the thermal optical timescale from the accretion disk of normal quasars. Four sources (NGC 7213, NGC 4051, Mkn 766, and NGC 3227) deviate from this trend, with timescales roughly consistent with the thermal optical regime.

We assign a physical meaning to the DRW mathematical model to understand these results, especially the timescales derived in DRW PSDs.
The steep slope (with spectrum index being -2) of the DRW PSD reflects the system's response to disturbances, where the state of the system undergoes significant changes due to the disturbance of injection of energy. Simultaneously, the system also experiences an energy-loss process that restores the system to equilibrium (with spectrum index being 0). 
The characteristic timescale corresponding to the spectrum index transformation can be considered as the relaxation/damping timescale, it is associated with the energy dissipation process.

Based on this physical scenario and our results, a natural explanation is that the coronal X-ray variability is driven by physical processes of the corona itself, which is different from the thermal optical variability in the accretion disk. 
The cooling or heating processes associated with the local thermal instability in the corona generally have relatively short timescales, which cannot account for the observed statistical timescales of 3-50 days.
This suggests that the variability with such timescales is unlikely to be caused by particle cooling or acceleration processes, but is more likely associated with the redistribution of thermal energy in the corona.
Thermal conduction is a relatively slow mechanism in the corona for energy regulation. 
It acts as a smoothing mechanism for temperature gradients across the corona.
We quantitatively estimate the conductive timescale (reflects the rate at which energy is redistributed due to temperature gradients) under reasonable values of physical parameters and investigate whether it can account for our observed X-ray timescale.
The expression of the conductive timescale is given by \citep{1999MNRAS.308..751R,2012ApJ...754...81L}:
\begin{equation}\label{t_cond}
\tau_{\rm cond}\sim\frac{u}{q/L}\sim \frac{\frac{3}{2}n_{e}k_{B}T}{-\kappa\frac{dT}{dL}/L}\sim \frac{\frac{3}{2}n_{e}k_{B}T}{\kappa\frac{T}{L^2}}= \frac{3n_{e}k_{B}L^2}{2f\kappa_{Sp}}\,s\;,
\end{equation} 
where $u$ is the thermal energy density, $q$ is the heat flux, $L$ is the coronal size. $T$ refers to the electron temperature which is typically in the range of $10^8-10^9$ K \citep{2022ApJ...927...42K} in a hot corona. $\kappa=f\kappa_{Sp}$ is the thermal conductivity in which $f$ is the suppression factor due to magnetic fields, and $\kappa_{Sp}=1.84\times 10^{-5}\;T^{5/2}\;\rm erg\;cm^{-1}\;s^{-1}\;K^{-1}$ is the Spitzer thermal conductivity \citep{1962pfig.book.....S}. $n_{e}\sim 10^{10}-10^{12}\;\rm cm^{-3}$ is the electron number density \citep{2018A&A...614A..37T,2022MNRAS.512..728C}, and $k_{B}$ is the Boltzmann constant.
Taking the typical values of these parameters, we tentatively estimate that conductive timescale could be consistent with the observed variability timescale of 3-50 days when $f$ and $L$ fall within reasonable ranges of 0.01-0.2 and $10^{12}-10^{14} \rm cm$ (equivalent to a few to tens of $R_{g}$ for AGN with black hole mass of $10^{7} M_{\rm \odot}$) respectively \citep{1998PhRvL..80.3077C,2001ApJ...562L.129N,2020MNRAS.495.1158C,2022MNRAS.512..728C,2025FrASS..1130392L}.
So, we propose that our observed X-ray variability may arise from a global thermal conduction process within the corona, where gradual redistribution of energy leads to changes in radiative output. 
The coronal X-ray timescales spanning an order of magnitude reflects possible differences in the physical conditions of the corona among these AGNs, such as coronal structure, magnetic field strength and particle density.

Our previous research demonstrates that the long-term variability in jet and accretion disk of AGNs are both related to the thermal instability of accretion disk \citep[Figure 15 in][]{2023ApJ...944..103Z}, because the variability timescales for jet and accretion disk are both consistent with the thermal instability timescale which is defined as 
\begin{equation}\label{t_instability}
t_{\rm th}=1680(\frac{\alpha}{0.01})^{-1}(\frac{M_{\rm BH}}{10^{8}M_{{\rm \odot}}})(\frac{R}{100R_{\rm S}})^{3/2}\;days, 
\end{equation} 
when the viscosity parameter $\alpha$ is set to be 0.05. $R$ represents the thermal emission distance from the central black hole. 
Considering disk-corona coupling, we propose that the coronal X-ray variability may also originate from the thermal instability of the accretion disk. 
Under the standard accretion disk theory, the coronal X-ray variability of NGC 7213, NGC 4051, Mkn 766 and NGC 3227 may raise from a similar region of the accretion disk where the thermal instability causes the variability in accretion disk (optical) and jet.
The coronal X-ray variability timescales of the remaining sources can also align with the predicted thermal instability timescale of the accretion disk at smaller radii ($R \simeq 20-40 R_{S}$), closer to the central black hole. 
This suggests that local thermal instabilities in the accretion disk (near the black hole), can efficiently propagate to the corona via disk-corona coupling, where the corona responds rapidly, producing X-ray variability on the corresponding timescales.
That is, under the assumption of the same standard thin disk, the thermal instability of the accretion disk may contribute to not only the variability in the accretion disk and the jet, but also the coronal X-ray variability. 
This may provide observational evidence of a correlation between the radiation of the fundamental material structures in AGN, i.e., the accretion disk, jet and the corona.

A correlation analysis between the optical/UV and X-ray light curves of AGN can serve as a further investigation to provide valuable insight into the origin of coronal X-ray variability.
It has been found that there exists a moderate correlation between the optical/UV  and X-ray variability, with the X-ray typically leading the optical/UV emission by several hours to a few days. Furthermore, the measured time lags tend to increase with wavelength, following a relation of $\tau \propto \lambda^{4/3}$ \citep[e.g.,][]{2015ApJ...806..129E,2017MNRAS.464.3194B,2018MNRAS.480.2881M,2021ApJ...922..151K,2024A&A...691A..60P}. 
These observations favor the X-ray reprocessing picture, in which coronal X-rays illuminate the disc and are thermally reprocessed into optical/UV bands.
Meanwhile, the observed behavior matches our thermal-conduction corona model, which provides a natural explanation for the origin of the X-ray variability.
Conversely, when the optical/UV bands lead or exhibit more intricate correlations with the X-rays, the variability could be primarily driven by fluctuations propagating inward from the accretion disc \citep{2020MNRAS.494.4057P,2024MNRAS.530.4850H}.
Discriminating between this propagation scenario and the X-ray reprocessing demands a more detailed analysis of the correlations between the optical/UV and X-ray variability for each AGN.

\section{Summary} \label{sec:summary}
We have applied the GP method to coronal X-ray (2-10 keV) LCs of 13 AGNs to investigate the variability properties in the corona and explore the connection between the radiation emitted from the basic material structures of AGNs, i.e., jet, accretion disk and the corona.
A lower range of black hole mass (compared to that of blazars in our previous studies) is extended in $\rm \tau_{damping}^{rest}-M_{BH}$ plot.
LCs for all our sample can be successfully described by DRW model, and the derived X-ray characteristic timescales are in the range of 3-50 days, most occupying a lower region compared to the optical timescales extracted from normal quasars in the plot of $\rm \tau_{damping}^{rest}-M_{BH}$.
We propose that the coronal X-ray variability may be driven by internal processes within the corona itself (e.g., thermal conduction), and also may be triggered by thermal instabilities of the accretion disk (close to the central black hole), which propagate to the corona via disk-corona coupling.

\acknowledgments
We thank the anonymous reviewers for constructive suggestions.
We thank the funding support from the Postdoctoral Fellowship Program of China Postdoctoral Science Foundation (CPSF) under Grant Number GZB20230618 and the National Natural Science Foundation of China (NSFC) under grant No. 12393852.

This work has made use of {lightcurves}{spectral files} provided by the University of California, San Diego Center for Astrophysics and Space Sciences, X-ray Group (R.E. Rothschild, A.G. Markowitz, E.S. Rivers, and B.A. McKim), obtained at \url{http://cass.ucsd.edu/~rxteagn/}.

{\it Facility:} RXTE 

{\it Software:} celerite \citep{2017AJ....154..220F}, emcee \citep{2013PASP..125..306F}, NumPy \citep{2020NumPy-Array}, Matplotlib \citep{2007CSE.....9...90H}, Astropy \citep{2013A&A...558A..33A,2018AJ....156..123A}, SciPy \citep{2020SciPy-NMeth}.

\bibliography{ms.bib}{}
\bibliographystyle{aasjournal}

\end{CJK*}
\end{document}